# Propagation Models for IEEE 802.15.6 Standardization of Implant Communication in Body Area Networks

*Raúl Chávez-Santiago, Oslo University Hospital*
*Kamran Sayrafian-Pour, NIST*
*Ali Khaleghi, K. N. Toosi University of Technology*
*Kenichi Takizawa, NICT*
*Jianqing Wang, Nagoya Institute of Technology*
*Ilangko Balasingham, Oslo University Hospital and Norwegian University of Science and Technology*
*Huan-Bang Li, NICT*

## Abstract

A body area network is a radio communication protocol for short-range, low-power, and highly reliable wireless communication for use on the surface, inside, or in the peripheral proximity of the human body. Combined with various biomedical sensors, BANs enable real-time collection and monitoring of physiological signals. Therefore, it is regarded as an important technology for the treatment and prevention of chronic diseases, and health monitoring of the elderly. The IEEE 802 LAN/MAN Standards Committee approved Task Group TG15.6 in December 2007. As a result of more than four years of effort, in February 2012, TG15.6 published the first international standard for BANs, IEEE Std 802.15.6. Throughout the development of this standard, ample collaboration between the standardization group and the research community was required. In particular, understanding the radio propagation mechanisms for BANs demanded the most research effort. Technical challenges were magnified for the case of implant communication because of the impossibility of conducting in-body measurements with human subjects. Therefore, research in this field had to make use of intricate computer simulations. This article outlines some of the research that has been done to obtain accurate propagation models supporting the standardization of implant communication in BANs. Current research to enhance the channel models of IEEE Std 802.15.6 through the use of ultra wideband signals for implantable devices along with physical measurements in animals is also presented.

## Introduction

The number of people with chronic diseases is constantly increasing worldwide. For the purposes of improving quality of life, efficient healthcare management, as well as disease prevention, real-time monitoring of various physiological signals is of paramount importance. Information such as electrocardiogram (ECG), body temperature, blood pressure, and endoscopic video can provide useful clues for doctors to diagnose whether a person is possibly threatened by any disease. That enables early and specialized pre-hospital management of patients. IEEE Std 802.15.6-2012 is a standard for short-range wireless communications to connect small devices deployed on the surface, inside or in the peripheral proximity of the human body [1]. Combined with various biomedical sensors, real-time measurement and monitoring of physiological signals becomes possible through the use of this technology. IEEE Std 802.15.6 specifies the characteristics of the physical (PHY) and medium access control (MAC) layers for the implementation of body area networks (BANs) in existing industrial, scientific, and medical (ISM) bands as well as frequency bands approved by national medical and/or regulatory authorities. The radio propagation channels of communication links in a BAN exhibit different characteristics [2]. For example, communication links among wearable nodes are prone to suffer from multipath fading and shadowing, whereas the transmitted signals from implant nodes undergo significant attenuation as they propagate through various tissues and organs.

In this article, we focus our attention on the implant communication in BANs only. Implantable devices offer myriad new and exciting medical applications such as implantable pills for precise targeted drug delivery, glucose monitors, bladder pressure monitors, smart capsule endoscopes, and micro robots operating inside the body for biopsy and therapeutic procedures [3]. However, since the early stage of the BAN standardization process, it was recognized that modeling the radio channel for implant communication would be a great challenge. Because of a number of ethical and technical issues, measurement campaigns inside living human bodies are not possible. Hence, in order to characterize the propagation of radio signals through human tissues, researchers must first carry out numerical simulations of electromagnetic waves propagation using a digital anatomical model. Then the obtained data can be processed statistically to produce simulation-based models. This article outlines some of the research that has been done to obtain accurate propagation models supporting the standardization of implant communication in BANs. Also, current research to enhance the channel models of IEEE 802.15.6 through the use of ultra wideband (UWB) signals for implantable devices along with physical measurements in animals is presented.

The rest of this article is organized as follows. In the next section, we chronologically describe the milestones of Task Group (TG) 15.6 and highlight the importance of radio propagation modeling in the standardization process. We summarize the research on implant channel models that supported the development of the standard. We present current research in the field of UWB propagation for implant communication. The results of this research may be added to the current standard in the future in order to widen its applicability. Perspectives for future collaboration between the research community and standardization groups toward the enhancement of the existing standard are described. Finally, concluding remarks are provided.

## The Standardization Process and Importance of Channel Models

IEEE 802 LAN/MAN Standardization Committee (LMSC) is an international organization that develops standards for wireless networks. Working Group 15 (WG15) under IEEE 802 focuses on wireless personal area networks (WPANs). TG15.6 was formally approved in December 2007 under WG802.15 with the objective of developing a standard for BANs.

The main milestones of TG15.6 are summarized in Table 1. After the start of TG15.6, four primary documents, used as guidelines for preparing proposals, were drafted and completed by November 2008. The four documents were:

- Regulation subcommittee report
- Application matrix
- Channel modeling
- Technical requirement

Then a call for proposals (CFP) was issued in November 2008. A total of 34 proposals were submitted and presented in March and May meetings of 2009. It took about 10 months for TG15.6 members to work out a common baseline, which was used as a guideline to merge the proposals. Then the editorial team worked quickly, and the first version of IEEE 802.15.6 (draft 01) was completed in July 2010. The draft was updated from version 01 to 06 during five rounds of letter ballots and four rounds of sponsor ballots. Finally, the standard was published in February 2012 as IEEE Std 802.15.6.

| Time stamp | Steps/achievements |
|---|---|
| December 2007 | Setup of TG15.6 |
| November 2008 | Completion of primary TG documents and issuing call for proposals |
| March–May 2009 | Submission and presentation of 34 proposals |
| March 2010 | Agreement on baseline and compromise on merge of 34 proposals |
| May 2010–July 2011 | Completion of draft 01-04 and letter ballot 01-05 |
| August–December 2011 | Completion of draft 05-06 and sponsor ballot 01-04 |
| February 2012 | Standard published as IEEE Std 802.15.6 |

**Table 1.** *Main steps and achievements in TG15.6.*

### The Need for a BAN Channel Model

In the above standardization process, common channel models were needed to fairly evaluate and compare the performance of different PHY proposals. The channel models had to characterize the path loss suffered by BAN devices taking into account shadowing caused by the different postures of the human body or objects in the vicinity. A large number of contributions to the channel models from academia, industry, and other research organizations were submitted and discussed during TG15.6 meetings. The final document that summarized the adopted BAN channel models was submitted in November 2010 [4].[1]

### Different BAN Propagation Scenarios

The Channel Modeling Subgroup within TG15.6 identified seven different propagation scenarios (S1–S7) in which IEEE 802.15.6 compliant devices may operate (Table 2). These scenarios are determined based on the locations of BAN nodes:

- Implant: inside the human body
- Body surface: in direct contact with the skin or within 2 cm distance
- External: beyond 2 cm and up to 5 m from the body surface

Scenarios S1, S2, and S3 correspond to implant nodes in a BAN. Two different channel models, CM1 and CM2, can be used to characterize the propagation scenarios for implant nodes. The frequency band 402–405 MHz has been allocated for medical implant communication services (MICS) by many international regulatory organizations including the Federal Communications Commission (FCC). A number of ultra-low-power medical implantable devices such as cardiac pacemakers and defibrillators already operate in

[1] *All the documents submitted to TG15.6 including the different revisions of the channel model can be found at https://mentor.ieee.org/802.15/documents*

Inhomogeneity and complexity of the propagation medium along with possible propagation paths from any direction necessitate a 3D environment to better capture, visualize, and understand RF propagation from/to implants.

| Scenario | Description | Frequency band | Channel model |
|---|---|---|---|
| S1 | Implant to implant | 402–405 MHz | CM1 |
| S2 | Implant to body surface | 402–405 MHz | CM2 |
| S3 | Implant to external | 402–405 MHz | CM2 |
| S4 | Body surface to body surface (LOS) | 13.5, 50, 400, 600, 900 MHz 2.4, 3.1–10.6 GHz | CM3 |
| S5 | Body surface to body surface (NLOS) | 13.5, 50, 400, 600, 900 MHz 2.4, 3.1–10.6 GHz | CM3 |
| S6 | Body surface to external (LOS) | 900 MHz 2.4, 3.1–10.6 GHz | CM3 |
| S7 | Body surface to external (NLOS) | 900 MHz 2.4, 3.1–10.6 GHz | CM3 |

**Table 2.** *Propagation scenarios for BANs [5].*

this frequency band. The MICS band offers good propagation behavior through human tissues and enables the use of reasonable-sized antennas, but its limited bandwidth constrains the communication devices to low data transmission rates.

### Path Loss Modeling

The path loss in free space between a transmitter and a receiver as a function of the distance separating them can be expressed by the Friis transmission equation [4]. The Channel Modeling Subgroup determined that this equation can also be used to model radio propagation in some of the BAN scenarios by adding a term to represent the random variations around the mean path loss value. Those variations are known as shadowing and are caused by the movement of the body or surrounding objects, and, in the case of implant channel, by varying dielectric properties of the different organs (heart, lung, liver, etc.) along the propagation path.

The TG15.6 relied on the propagation models CM1 and CM2 for the standardization of implant communication in BANs. Together with CM3, the TG15.6 was able to evaluate the PHYs within the 34 BAN proposals. The development of these propagation models, however, was achieved through highly innovative research. As physical radio channel measurements in human subjects are practically impossible, sophisticated simulation tools were the only option available to investigate the electromagnetic wave propagation inside the human body. The National Institute of Standards and Technology (NIST) in the United States facilitated the development of the standard for implant communication through a contribution submitted in September 2008. The contribution outlined a statistical path loss model for MICS [5].

## A Path Loss Model for MICS

### A 3D Immersive Visualization and Simulation Platform

Inhomogeneity and complexity of the propagation medium along with possible propagation paths from any direction necessitate a 3D environment to better capture, visualize, and understand radio frequency (RF) propagation from/to implants. Immersive platforms have been used in the past to gain better understanding of the physical phenomena occurring during an experiment. The complexity of the propagation environment surrounding a medical implant would also imply that an appropriately designed immersive platform would be very helpful in better understanding the RF channel characteristics inside the human body. Therefore, an innovative 3D virtual reality simulation platform aimed at RF propagation study in BANs was developed at the Information Technology Laboratory of NIST to accomplish this task [6].

The platform has four main components, as shown in Fig. 1a: a 3D immersive platform, a 3D human body model, a 3D full wave electromagnetic simulator, and a BAN antenna. The 3D immersive platform shown in Fig. 1b includes several important components: three orthogonal screens that provide the visual display, the motion tracked stereoscope glasses, and the handheld motion tracked input device. The screens are large projection video displays that are placed edge-to-edge in a corner configuration. These three screens are used to display a single 3D stereo scene. The scene is updated based on the position of the user as determined by the motion tracker. This allows the system to present to the user a 3D virtual world within which the user can move and interact with the virtual objects as depicted in Fig. 1b. The 3D human body model includes frequency-dependent dielectric properties of more than 300 parts in a male human body. These properties are also user-definable if custom changes or modifications are required. The propagation engine allows computing a variety of different electromagnetic quantities such as the magnitude of electric and magnetic fields or specific absorption rate (SAR).

The final component of the system is the implantable antenna. The operating environment for an implantable antenna is quite differ-ent from traditional free space communication. Designing an efficient antenna for implantable devices is an essential requirement for reliable network operation. The dimension of the antenna must be very small and should be long-term biocompatible.

Input parameters to the above virtual reality system include: a virtual antenna including all relevant parameters (i.e., position, orientation, directionality, polarization, etc.), operating frequency, transmit power, body model resolution, range, and choice of the desired output parameters. Using this environment, a researcher is able to place an implant antenna at the desired location of the human body and study the RF propagation at the target frequency. The platform user can look at data representations at any scale and position, physically move through data, change orientation, and control the elements of the virtual world using a variety of interactive measurement and analysis techniques. All of these capabilities are extremely useful when studying RF propagation to/from implanted devices. This platform also allows more natural interaction between experts with different backgrounds. For example, using the interactive tool, a surgeon can point to the exact location of the implant, and then, given the physical and biological constraints, an antenna designer can design an antenna to fit the chosen location. Finally, an RF engineer would be able to study the propagation performance for the desired communication link.

## A Statistical Path Loss Model for the MICS Channel

To obtain a statistical path loss model for the MICS band in IEEE 802.15.6, the virtual reality platform needed to be configured using appropriate input parameters. For example, the operating frequency was chosen to be 403.5 MHz, which is the mid-point of the MICS frequency band. The implant antenna used in the study is a multi-turn loop antenna (see a detailed description in [7]) composed of a single metallic layer that is printed on a side of a D51 (NTK) substrate with thickness of 1 mm. The dimension of the antenna is $8.2 \times 8.1 \times 1$ mm, which is quite appropriate for some medical applications.

After configuring the system, simulations were performed for four near-surface and two deep-tissue implant scenarios. The scenarios were carefully chosen to match real-life applications. For instance, the near-surface scenarios included applications such as an implantable cardioverter-defibrillator (ICD) and pacemaker (below the left pectoral muscle), Vegus nerve stimulation (right neck and shoulder), and two motion sensor applications (right hand and leg). The deep tissue implant scenarios addressed video capsule endoscopy applications for the upper and lower stomach. For each scenario (i.e., transmitter antenna location), the received power was calculated for a grid of points within a cylinder area around the body. Then the resulting data was partitioned into three sets: in-body to in-body, in-body to body surface, and in-body to out-of-body propagation sets. The in-body to body surface set included all points that reside within a user definable distance (i.e., 2 mm) from the body surface.

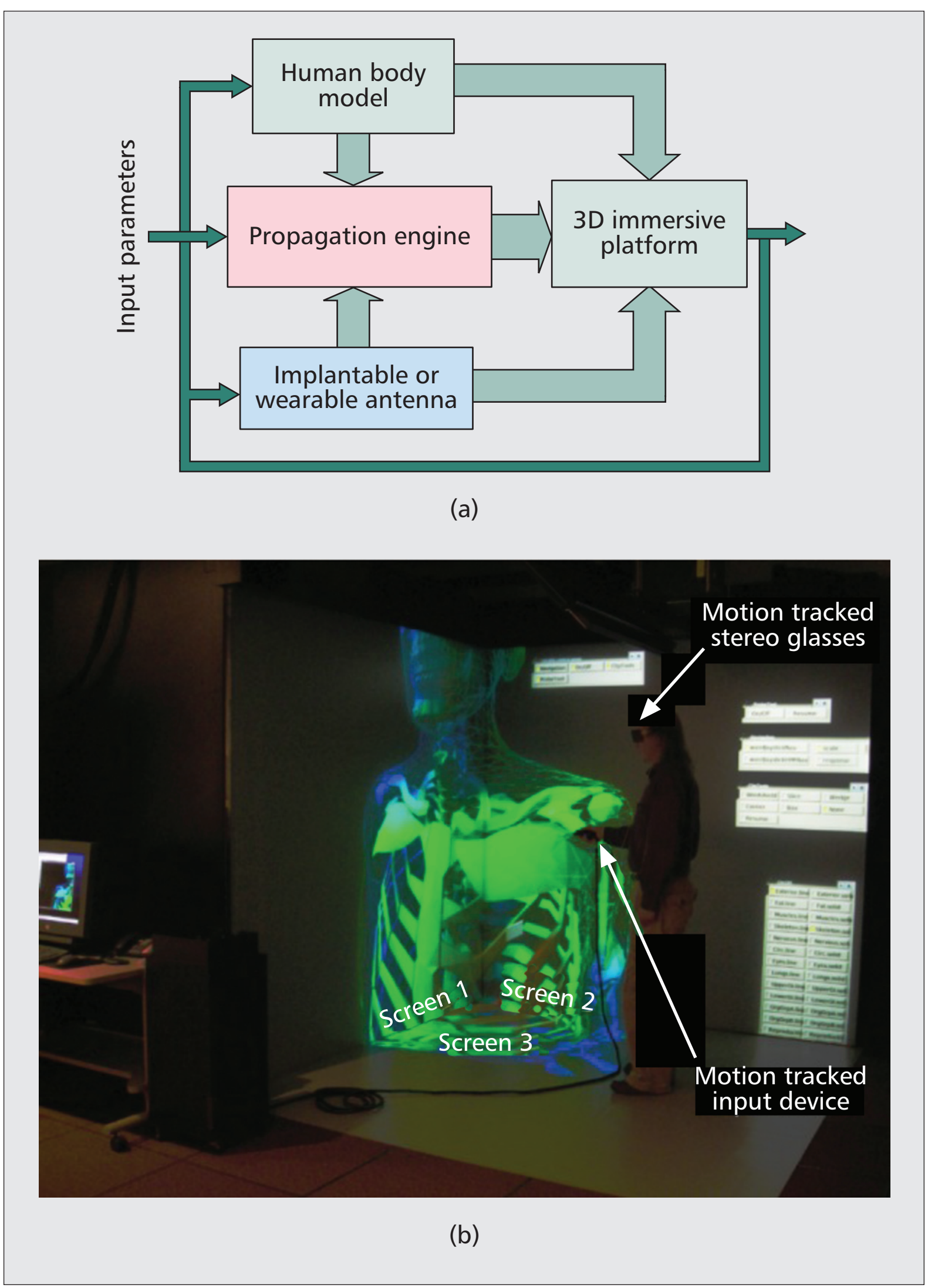

**Figure 1.** *The 3D immersive visualization and simulation platform developed at NIST [6]: a) system block diagram; b) physical implementation.*

Figure 2a shows the scatter plot for the path loss as a function of the transmitter-receiver separation for deep tissue implant to body surface scenarios. The mean value of the random path loss has been displayed by a solid line. This is obtained by fitting a least squares linear regression line through the scatter of measured path loss sample points in dB such that the root mean square deviation of sample points about the regression line is minimized. Random shadowing and scattering effects of the channel (due to inhomogeneity of the environment) occur where the transmitter-receiver separation is the same, but have different directions or positions with respect to each other. As shown in Fig. 2b, this term can be modeled by a normal distribution with zero mean and standard deviation of 7.85.

The 3D virtual reality platform is a powerful tool that enables extensive research in this cross-disciplinary field of engineering [8]. For example, studying RF propagation for specific medical implant applications such as cochlear implants, brain neural interface, and retina implants; and performance evaluation of custom made antennas including the surrounding environment (e.g., textiles, other objects) is possible using this platform. Also, considering that issues such as various body types, postures, and motion could further impact the propagation channel, research is underway to add motion capability to the 3D human model in order to study dynamic implant channel characteristics.

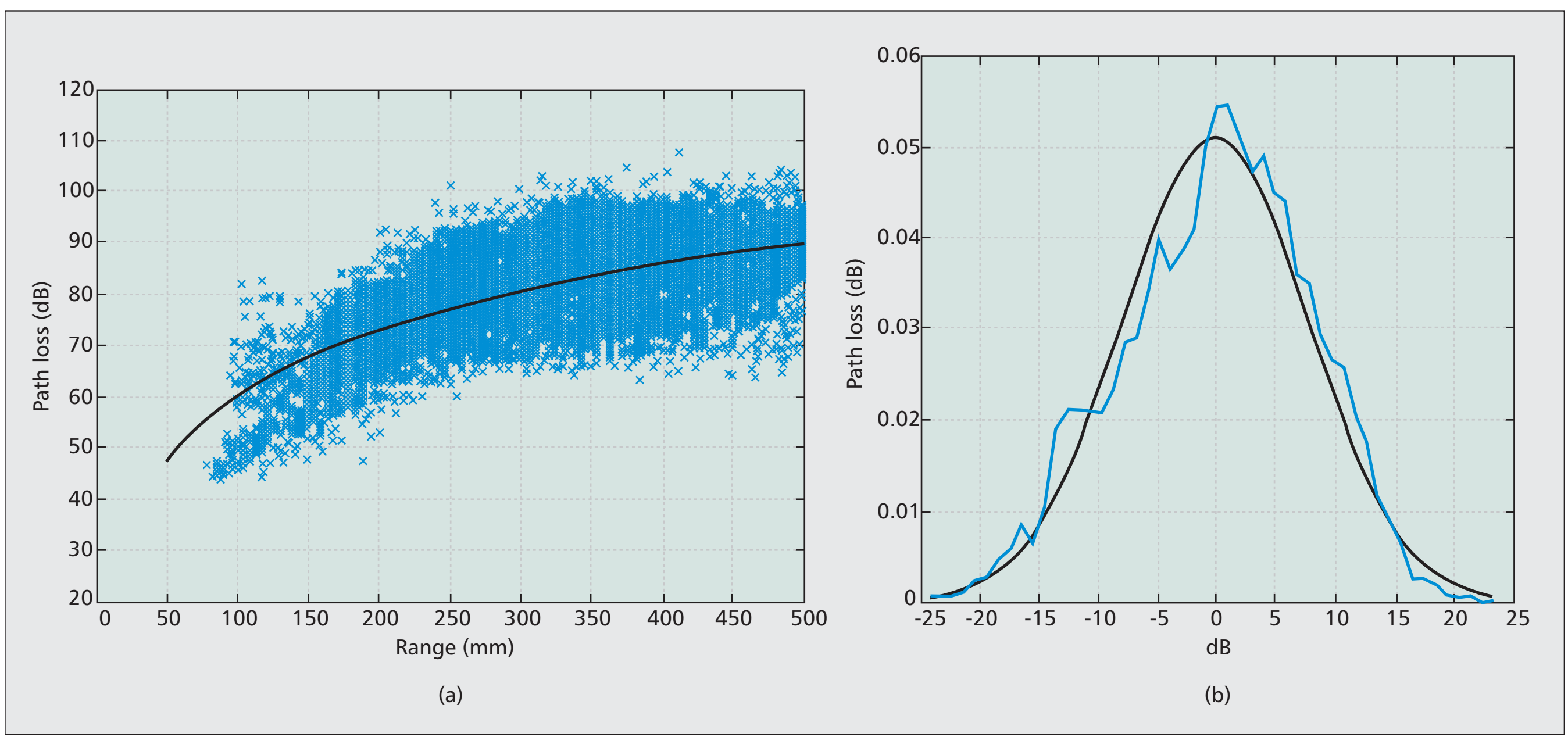

**Figure 2.** *Deep tissue implant to body surface [6]; a) scatter plot of the path loss vs. distance; b) distribution of the shadow fading.*

## Current Research on Propagation for Implant Communication

### Ultra Wideband Technology

Ultra wideband technology shows great potential for use in BAN applications. Aside from the available bandwidth capabilities of at least 500 MHz, UWB signals have other attractive characteristics. For example, they exhibit an inherent noise-like behavior due to their extremely low maximum effective isotropic radiated power (EIRP) spectral density of –41.3 dBm/MHz. This makes UWB signals difficult to detect and increases their robustness against jamming, thereby alleviating the need for complex encryption algorithms in small low-cost transceivers. Additionally, UWB transmitters are not significant sources of interference to other radio devices and do not represent a threat to human safety [9]. In particular, impulse radio (IR) UWB transceivers have a simple structure and very low power consumption. This facilitates their miniaturization for use in BANs. The Channel Modeling Subgroup at TG15.6 has already developed a UWB propagation model for wearable BAN nodes, which is part of CM3 for scenarios S4–S7 (Table 2). In [9], a simulation-based study demonstrated that high-data-rate communication with implantable devices (scenario S2) is also feasible with IR-UWB transceivers. This was a case study for the design of an IR-UWB communication system for a capsule endoscope operating in the 1–5 GHz frequency range, that is, the lower part of the FCC allocated UWB frequency band (3.1–10.6 GHz). High-data-rate communication with implantable sensors provides new opportunities for innovative healthcare applications, as mentioned before; however, further understanding of the UWB implant channel is required. This is a challenging task due to the highly complex nature of the propagation medium.

### Ultra Wideband Propagation Models

There have been limited attempts to model the propagation channel for UWB implantable devices by the research community. This is mainly due to the high cost of the necessary digital anatomical models and simulation tools. In fact, to the best of our knowledge only two models of UWB propagation for implantable devices in the human chest have been reported [10, 11].

The first one (i.e., [10]) was developed at the Nagoya Institute of Technology (NIT), Japan, through numerical simulations using a voxel anatomical model that includes nearly 50 types of tissue with a spatial resolution of 2 mm. This voxel model of an adult male was developed at the National Institute of Information and Communications Technology (NICT) in Japan based on magnetic resonance imaging (MRI) data. To collect channel data, an implantable disc dipole antenna was designed to cover the frequency range of 3.4–4.8 GHz. This range corresponds to the low UWB band allowable for transmission in Japan. Two of these antennas were embedded in the anatomical model, one as a transmitter inside the body and the other as a receiver on the chest. A 2nd-derivative Gaussian pulse with duration of 2.1 ns was applied to the implanted antenna, which was placed inside the left chest at 20 different arbitrarily chosen locations ranging from 6 to 20 mm from the skin. The on-body receiver was located in an area of 100 × 100 mm in the front of the left chest with a distance up to 200 mm from the chest surface. In this area, nine arbitrary receiver locations were considered. As a result, 180 transmitter-receiver links were simulated using the finite-difference time-domain (FDTD) method. Using these data, a channel model for the channel impulse response (CIR) of the in-body to out-of- body UWB link (scenario S3) at the chest area was obtained. The model predicts a root mean square (RMS) delay spread of around 0.2 ns.

The second model [11] was obtained at the Intervention Centre, Oslo University Hospital, Norway. It includes a statistical description of both the CIR and path loss of a UWB link (scenario S2) for the human chest. The model was developed based on the anatomical data set of the Visible Human Project®, which is a voxel representation of an adult male. This anatomical model includes 24 different tissues with a spatial resolution of 2 mm. The anatomical model was embedded in a simulator that uses the time-domain finite integration technique (FIT). Unlike the approach used in [10], a transmitting antenna was not used in the simulation scenario. Instead, the chest was exposed to an incoming plane wave from the front. Ideal electric field probes were placed inside the chest. They were arranged in seven planes consisting of 85 probes each and parallel to the skin for a total of 595 probes. The plane wave was excited with a 2nd-derivative Gaussian pulse with duration of 420 ps in order to have most of the pulse energy within 1–6 GHz. The received signal strength at every probe location was obtained and plotted vs. distance from the skin (Fig. 3). This model predicts an RMS delay spread below 1 ns, which is in agreement with the results in [10]. Moreover, the scattering around the mean path loss increases with the distance. This is due to higher tissue non-homogeneity as depth increases. By including the frequency-selective attenuation of the channel in the path loss computation, further improvement to this model has been made [12]. In UWB implant communication, the high frequency-selective attenuation reduces the effective bandwidth of a transmitted signal as it propagates through living tissues. The development of an accurate UWB path loss model for implant communication requires thorough investigation of this frequency selectivity.

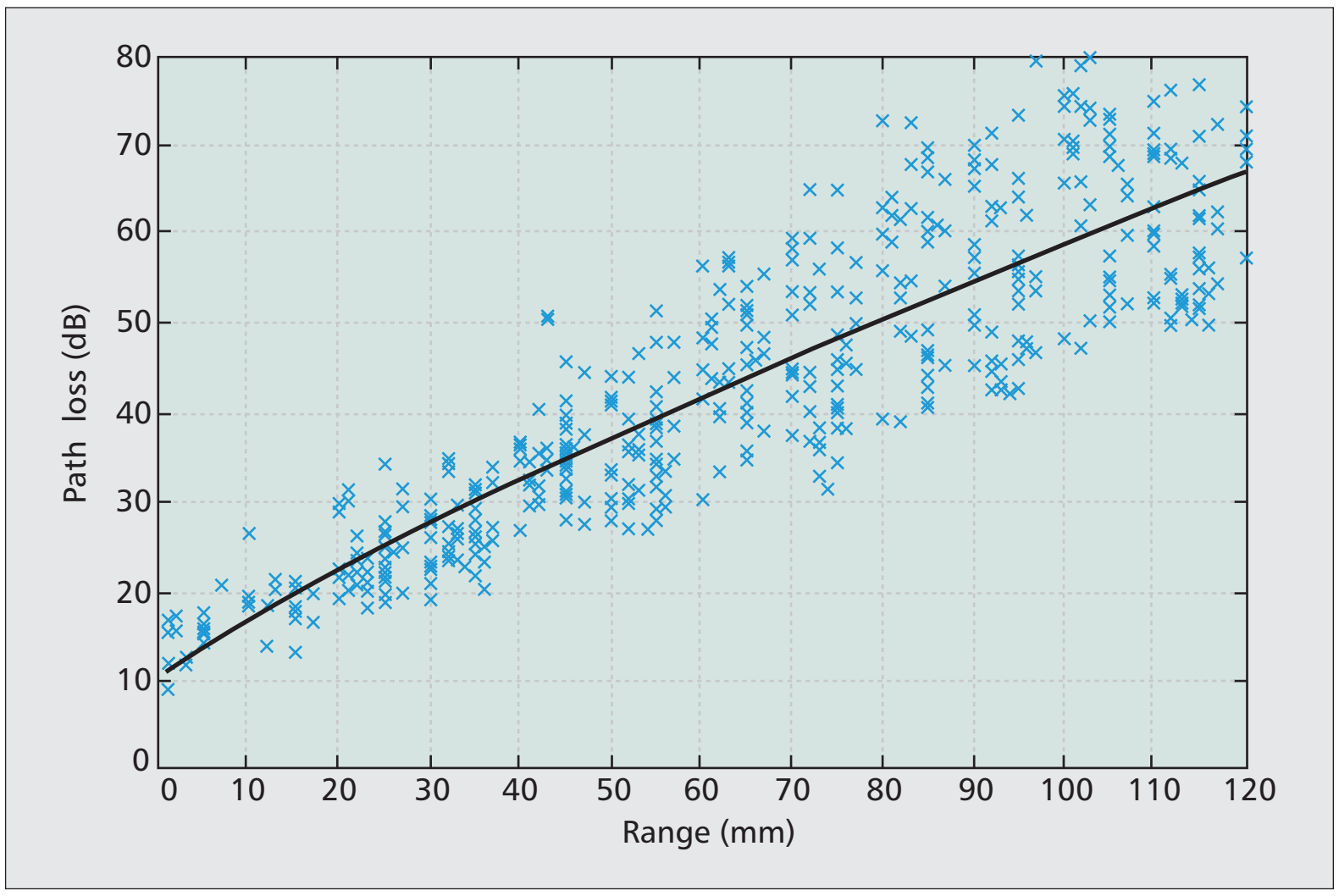

**Figure 3.** *Scatter plot of the path loss in the 1–6 GHz band versus distance for implant to body surface [11].*

### The Need for a Common Simulation Platform

The use of different anatomical models and simulation scenarios for the development of statistical propagation models, in addition to many other factors that impact the outcome of the numerical computations, can lead to discrepancies between the predicted path loss values. It has been debated whether using a plane wave or a transmitting antenna is more convenient for modeling UWB radio propagation for scenarios S2 and S3. Strictly speaking, the radio channel consists of the propagation medium and the transmitting/receiving antennas. But different antenna structures lead to different radio channels; therefore, using a plane wave allows for the characterization of the propagation medium individually. However, the drawback of the plane wave approach is the inability to properly model the near-field coupling effects for short communication links. Researchers must reach a consensus on the simulation platform and technique to develop a common propagation model for the standardization of UWB-based implant communication.

Regardless of the selected simulation approach, the validation of the statistical model is still a problem that must be solved. The best way of validating a propagation model is through physical measurements in a realistic propagation medium. However, in-body experimentation in humans is practically impossible. Therefore, an alternative for validation of a particular statistical model could be through comparisons with other models developed independently by different research groups, or through measurements in specially formulated chemical solutions that mimic the average dielectric properties of human biological tissues (phantoms) [13]. However, since other statistical models are based on simulation too, there will still be uncertainty in their accuracy and reliability. Another alternative for validating a model could be using a theoretical model developed from the fundamental principles of electromagnetism; but such a mathematical model is very difficult to produce due to the extreme complexity of the propagation medium. Finally, measurements in phantoms do not capture the complex propagation phenomena observed in inhomogeneous layers of materials with different dielectric constants such as the human body.

## Perspectives for the Improvement of the Standard

### Radio Channel Measurements in Animals

A strategy for the validation of the existing MICS and future UWB models for the implant communication channel (as outlined by the TG15.6 leadership) is expansion of research to involve expertise from the healthcare domain. Through such collaborations, the validation of statistical propagation models can be done by

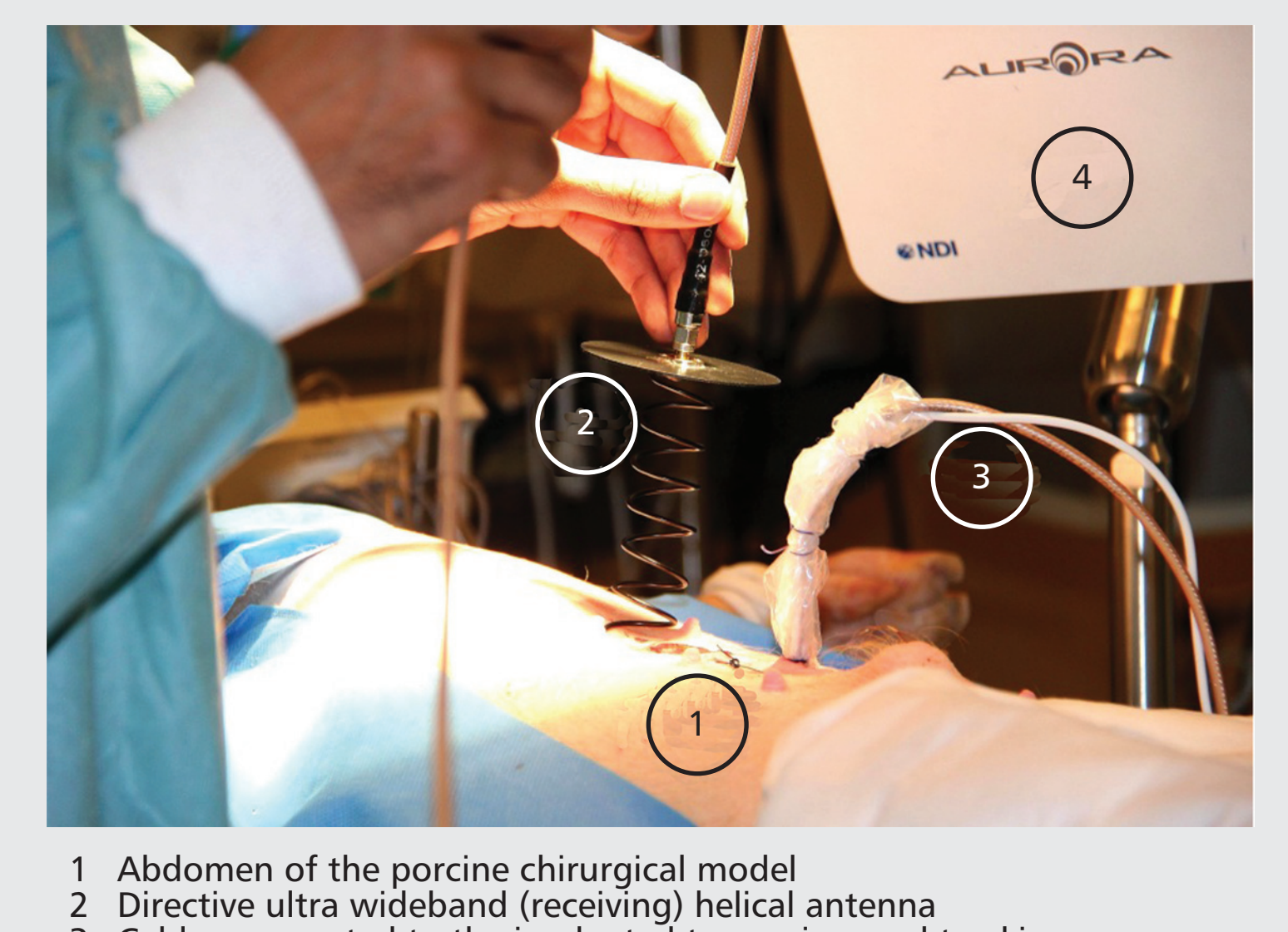

1 Abdomen of the porcine chirurgical model
2 Directive ultra wideband (receiving) helical antenna
3 Cables connected to the implanted transceiver and tracking sensor
4 Magnetic field generator for the tracking system performing experiments on living animals that anatomically resemble parts of the human body. Recently, measurements of the UWB propagation channel in a porcine model have been done at Oslo University Hospital (Fig. 4). NICT in Japan, a key player in the IEEE 802.15.6 standardization process, provided the UWB transceiver in 4224–4752 MHz that was implanted in the abdomen of the animal following the clinical procedure approved by the Research Animal Commission of Norway. Since a flat spectrum shape simplifies the measurements, multiband orthogonal frequency-division multiplexing (MB-OFDM) UWB was used. A highly directive UWB helical antenna on the animal's skin received the signal from the transmitter located inside the abdomen. A magnetic tracker system measured the distance between the implanted device and the skin with a margin of error of ±0.7 mm. Preliminary results showed an attenuation of 38.5 dB for a depth of 27 mm. The mean path loss for the same depth predicted by the model in Fig. 3 is 28 dB. This discrepancy appears because the statistical model does not consider any antenna distortion or cable attenuation. Hence, more measurements are required to obtain an offset value that accounts for additional losses not captured by the computer simulations. Moreover, enough data must be collected to ensure the statistical significance of the validation. Such measurement campaign is currently underway. The MICS model [4] is also expected to be validated through similar channel measurements in the near future.

**Figure 4.** *Measurement of the signal level from an implanted UWB transceiver in a porcine chirurgical model.*

### Body Area Networks with Mobile Implants

Novel applications in medicine such as capsule endoscopes require the inclusion of mobility in the propagation model for scenario S2. Although in-body mobile devices travel rather slowly (a capsule endoscope takes an average of 8 h to traverse the digestive tract), their continuous movement could cause wide variations of the received power level. The low UWB band (3.1–4.8 GHz) is seen as a good candidate for implementing the communication link of these high-data-rate in-body mobile applications because of the accurate localization and tracking capability of UWB signals. Some research advances have been made for the development of an accurate channel model tailored to UWB communication with in-body mobile devices. In [14] a UWB channel model for a capsule endoscope was presented. This statistical model was developed with the same anatomical model as in [11]. The path loss was given for several receivers located on a belt around the waist, and it was modeled by a Fourier double series (see [14] for mathematical details). In [15] a spatial diversity reception technique for a capsule endoscope in 3.4–4.8 GHz was proposed. This study was done with the same anatomical model as in [10]. Theoretical and computer simulation results demonstrate that spatial diversity effectively counters the large attenuation observed in a UWB link for mobile implants. With an appropriate arrangement of diversity receivers on the body surface, an improvement of 10 dB on the $E_b/N_0$ at a bit error rate of $10^{-3}$ can be obtained.

### Compliance with the Requirements from the Biomedical Industry

Path loss is a necessary parameter for the design and evaluation of a communication link. The available time dispersion (CIR) models provide numerical insight on path loss in relation to transmission range, frequency, and multipath effect. Thus, they can be used for either design or evaluation of communication links including error rate and throughput performance. The performance criteria of a communication scheme depend on the transmission content. For most types of data, the desired bit error rate is usually considered to be between $10^{-6}$ and $10^{-9}$. Using forward error correction (FEC) techniques, the tolerable packet error rate could be much higher. In IEEE 802.15.6, this rate has been set to at most 10 percent for a 256-octet payload with a link success probability of 95 percent [16].

## Conclusions

The development of the IEEE 802.15.6TM-2012 standard is a perfect example of successful collaboration between academia, research institutions, and a standardization group. The synergy between research and standardization activities was more evident in the extraction of the channel models that served as the foundations for this new standard. However, there is still room for extension and improvement of the standard concerning implant communication in body area networks. The addition of ultra wideband to these scenarios could enable a new set of applications due to the high-data-rate transmission capability of this technology. Hence, collaborative work toward obtaining a new ultra wideband channel model for the standardization of the network of implantable devices is necessary. Pro-moting the multidisciplinary nature of the research activities is an important strategy that a standardization group such as TG15.6 should pursue. The participation of experts with clinical backgrounds has opened the possibility of performing radio channel measurements in living animals for the validation of statistical models obtained through computer simulations. This strategy will overcome the limitations of performing measurements in humans. Clearly, the collaboration between all participants has not ended with the release of the standard. More work for the improvement of the current standard is expected in the coming years.

## Acknowledgment

R. Chávez-Santiago, A. Khaleghi, and I. Balasingham acknowledge financial support from the Research Council of Norway, given through the MELODY Project (contract no. 187857/S10).

## Biographies

Raúl Chávez-Santiago (raul.chavez-santiago@rr-research.no) obtained his Ph.D. degree in electrical and computer engineering from Ben-Gurion University of the Negev, Israel, in 2007. He held postdoctoral positions at the Université Paris-Sud XI, France, and Bar-Ilan University, Israel. He joined the Intervention Centre, Oslo University Hospital, Norway, in 2009. He currently investigates short-range radio communication technology (including UWB) for implant medical applications. He is a Management Committee member in COST Actions IC0902 and IC0905 on cognitive radio.

Kamran Sayrafian-Pour [SM] (kamran.sayrafian@nist.gov) holds a Ph.D. in electrical and computer engineering from the University of Maryland. He is currently a program lead at the Information Technology Laboratory of the National Institute of Standards and Technology located in Maryland. He was a contributing member and co-editor of the channel modeling document of the IEEE802.15.6 international standardization on BANs. He is also an adjunct faculty member of the University of Maryland.

Ali Khaleghi (ali.khaleghi@rr-research.no) received his Ph.D. degree in physics from the University of Paris XI, France. He joined the Institut d'Electronique et de Télécommunications de Rennes (IETR), France, as a postdoctoral researcher working on time reversal wireless communications. He started a postdoctoral position at the Intervention Centre, Oslo University Hospital, Norway, in 2008. There, he worked on wave propagation inside and on the human body. Currently, he is an assistant professor at the K. N. Toosi University of Technology.

Kenichi Takizawa (takizawa@nict.go.jp) received his B.E., M.E., and Dr.Eng. degrees from Niigata University, Japan, in 1998, 2000, and 2003, respectively. He joined the Communications Research Laboratory, National Institute of Information and Communication Technology (NICT), Yokosuka, Japan, in 2003.

Jianqing Wang (wang@nitech.ac.jp) received his B.E. degree from Beijing Institute of Technology, China, in 1984, and M.E. and D.E. degrees from Tohoku University, Sendai, Japan, in 1988 and 1991, respectively. He was a research associate at Tohoku University and a senior engineer at Sophia Systems Co., Ltd., prior to joining Nagoya Institute of Technology, Japan, in 1997, where he is currently a professor. His research interests include biomedical communications and electromagnetic compatibility.

Ilangko Balasingham (ilangko.balasingham@medisin.uio.no) received M.Sc. and Ph.D. degrees from the Department of Telecommunications, Norwegian University of Science and Technology (NTNU), in 1993 and 1998, respectively. He heads the Wireless Sensor Network Research Group at Oslo University Hospital and is a professor of medical signal processing at NTNU. His research interests include robust short-range wireless communications and localization. He was General Chair of the 2012 BodyNets Conference and is Editor of the *Nano Communication Networks Journal*.

Huan-Bang Li (lee@nict.go.jp) received his B.S. degree from Northern Jiao Tong University, China, in 1986, and M.S. and D.E. degrees from Nagoya Institute of Technology in 1991 and 1994, respectively. Since then, he has been with CRL/NICT, Japan. From 1999 to 2000, he was a visiting scholar at Stanford University, California. He currently works on UWB and BANs. He is a visiting professor at the University of Electro-Communications, Japan.